\documentclass[a4paper,useAMS,usenatbib]{mnras}

\usepackage{graphicx}
\usepackage{setspace}
\usepackage{natbib}
\usepackage{color}
\usepackage{amsmath,amssymb}
\usepackage{times}
\usepackage{hyperref}

\newcommand{\Gaia}{{\it Gaia}}

\title[Co-formation of the Galactic disc and the stellar halo]{Co-formation of the disc and the stellar halo\thanks{This paper is dedicated to the memory of Professor Donald Lynden-Bell}}

\author[Vasily A. Belokurov et al]{V. Belokurov$^{1,2}$\thanks{E-mail:vasily@ast.cam.ac.uk},  D. Erkal$^{1,3}$, N.W. Evans$^1$, S.E. Koposov$^{1,4}$ and A.J. Deason$^{5}$\\
  $^{1}$Institute of Astronomy, Madingley Rd, Cambridge, CB3 0HA\\
  $^{2}$Center for Computational Astrophysics, Flatiron Institute, 162 5th Avenue, New York, NY 10010, USA\\
  $^3$Department of Physics, University of Surrey, Guildford GU2 7XH, UK\\
  $^4$Department of Physics, McWilliams Center for Cosmology, Carnegie Mellon University, 5000 Forbes Avenue, Pittsburgh, PA 15213, USA\\
  $^{5}$Institute for Computational Cosmology, Department of Physics, University of Durham, South Road, Durham DH1 3LE, UK
}

\begin{document}

\maketitle

\label{firstpage}

\begin{abstract}
Using a large sample of Main Sequence stars with 7-D measurements
supplied by \Gaia\ and SDSS, we study the kinematic properties of the
local (within $\sim$10 kpc from the Sun) stellar halo. We demonstrate
that the halo's velocity ellipsoid evolves strongly with
metallicity. At the low [Fe/H] end, the orbital anisotropy (the amount
of motion in the radial direction compared to the tangential one) is
mildly radial with $0.2<\beta<0.4$. However, for stars with
[Fe/H]$>-1.7$ we measure extreme values of $\beta\sim0.9$. Across the
metallicity range considered, i.e. $-3<$[Fe/H]$-1$, the stellar halo's
spin is minimal, at the level of $20<\bar{v}_{\theta}(\mathrm{kms}^{-1}) <30$. Using a suite of cosmological zoom-in simulations of halo formation,
we deduce that the observed acute anisotropy is inconsistent with the
continuous accretion of dwarf satellites. Instead, we argue, the
stellar debris in the inner halo were deposited in a major accretion
event by a satellite with $M_{\rm vir}>10^{10} M_{\odot}$ around the
epoch of the Galactic disc formation, i.e. between 8 and 11 Gyr
ago. The radical halo anisotropy is the result of the dramatic
radialisation of the massive progenitor's orbit, amplified by the
action of the growing disc.
\end{abstract}

\begin{keywords}
Milky Way -- galaxies: dwarf -- galaxies: structure -- Local Group -- stars
\end{keywords}

\section{Introduction}

Stars in the Galactic halo formed earlier compared to those in the
disc. However, the epoch of the halo assembly, i.e. the time at which
these stars were deposited into the Milky Way, remains blurry. The
onset of star-formation in the Galactic disc is bracketed to have
happened between $\sim$8 and $\sim$11 Gyr ago, based on white
dwarf cooling ages \citep[see][]{Oswalt1996, Leggett1998, Knox1999,
  Kilic2017}, isochrone modelling
\citep[e.g.][]{Haywood2013,Martig2016} and nucleocosmochronology
\citep[][]{delPeloso2005}.  The birth times of the halo's stellar
populations can be estimated following the same procedures as for the
disc. For example, it has been determined that the halo's globular
clusters range in age from $\sim$10 to $\sim$13 Gyr depending on
metallicity \citep[see
  e.g.][]{Hansen2002,Hansen2007,Vandenberg2013}. Most dwarf
spheroidals host stars nearly as old as the Universe itself
\citep[see][]{Tolstoy2009}, while many ultra-faint dwarfs appear to
have formed the bulk of their stellar populations only slightly less
than a Hubble time ago \citep[see e.g.][]{Belokurov2007,Brown2014}. In
the field, the halo stars have ages in the range of 10-12 Gyr
\citep[][]{Jofre2011,Kilic2012,Kalirai2012}.

The first attempt to decipher the history of the formation of the
Galactic disc and the stellar halo can be found in \citet{ELS}. Based
on the strong apparent correlation between the metallicity of stars in
the Solar neighborhood and the eccentricity of their orbits, the
authors conclude that the metal-deficient stars formed first in an
approximately spherical configuration. This original gas cloud then
collapsed in a nearly instantaneous fashion, giving birth to the
subsequent generations of stars. The contraction of the Galaxy not
only yielded a population of younger stars on nearly circular orbits,
but also profoundly affected the orbital shapes of the old stellar
halo. As \citet{ELS} demonstrate, driven by the collapse, the orbits
of the halo stars ought to become highly eccentric, thus explaining
the chemo-kinematic properties of their stellar sample. Later studies
\citep[see e.g.][]{Norris1985, Beers1995, Carney1996,Chiba1998}
revealed that the claimed correlation between the eccentricity and
metallicity might have been caused by a selection bias affecting the
dataset considered by \citet{ELS}. Notwithstanding the concerns
pertaining to the observational properties of their data, the
theoretical insights into the dynamics of the young Galaxy exposed by
\citet{ELS} are as illuminating today as they were 50 years ago.

Whilst the hypothesis of strict ordering of the orbital properties
with metallicity in the stellar halo is not supported by the data,
undeniably, a large fraction of the halo stars are metal-poor and do
move on eccentric orbits. In spherical polar coordinates one can
characterize the shape of the stellar halo's velocity ellipsoid by one
number, the anisotropy parameter: $
\beta=1-\frac{\sigma_{\theta}^2+\sigma_{\phi}^2}{2\sigma_r^2}$, where
$\beta=-\infty$ for circular orbits and $\beta=1$ for radial ones.  In
the Solar neighborhood, the halo's velocity ellipsoid is evidently
radially biased. For example, \citet{Chiba1998} obtained $\beta=0.52$
for a small sample of local halo red giants and RR Lyrae observed by
the Hipparcos space mission. Using the SDSS Stripe 82 proper motions,
and thus going deeper, \citet{Smith2009} measured $\beta=0.69$ using
$\sim2,000$ nearby sub-dwarfs. Combining the SDSS observations with
the digitized photographic plate measurements, \citet{Bond2010}
increased the stellar halo sample further and derived
$\beta=0.67$. For slightly larger volumes probed with somewhat more
luminous tracers, similar values of $\beta\sim0.5$ were obtained
\citep[see e.g.][]{Deason2012,Kafle2012}. Unfortunately, beyond 15-20
kpc from the Sun, the behaviour of the anisotropy parameter is yet to be
robustly determined. While there have been several attempts to tease $\beta$
out of the line-of-sight velocity measurements alone \citep[see
  e.g.][]{Sirko2004,Wi15}, \citet{Hattori2017} have shown that these
claims need to be taken with a pinch of salt. Curiously, the only
distant stellar halo anisotropy estimate which actually relies on the
proper motion measurements (with HST) reports a dramatic drop to
$\beta\sim0$ at $\sim$20 kpc
\citep[see][]{Deason2013a,Cunningham2016}, albeit based on a very
small number of stars.

\begin{figure}
  \centering
  \includegraphics[width=0.48\textwidth]{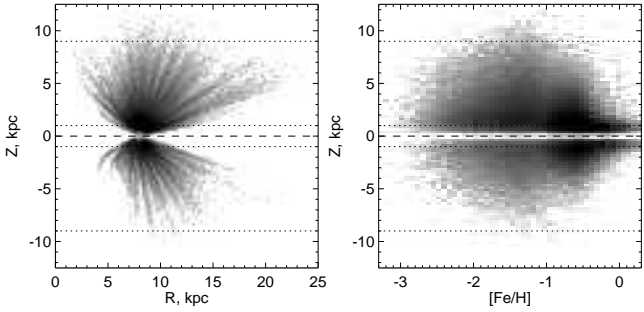}
  \caption[]{Galacto-centric view of the SDSS-\Gaia\ sample. {\it
      Left:} Logarithm of the stellar density in cylindrical $R,z$
    coordinates. {\it Right:} Logarithm of stellar density in the
    plane of [Fe/H] and Galactic height $z$. Horizontal dotted lines
    give the range of heights considered in the velocity ellipsoid
    analysis.}
   \label{fig:sample}
\end{figure}

Theoretically, the behavior of the orbital anisotropy in the dark halo
has been scrutinized thoroughly. As demonstrated by
\citet{Navarro2010}, in the {\it Aquarius} suite of simulations,
$\beta$ starts nearly isotropic in the centre of the galaxy, it
reaches $\beta\sim0.2$ around the Solar radius and keeps growing to
$\beta\sim0.5$ at $>$100 kpc, before falling again to zero around the
virial radius. The picture gets somewhat muddled when the effects of
the baryons are included. According to \citet{Debattista2008}, halo
contraction due to baryonic condensation may make the DM halos
slightly more radially anisotropic, but overall, the changes are small
and the $\beta$ trends with Galacto-centric distance are preserved. On
the other hand, when the {\it Aquarius} halos are re-simulated with
added gas physics and star-formation, only some galaxies retain their
DM-only anisotropy trends, while in others, the $\beta$ profile flattens and
stays constant with $\beta\sim0.2$ throughout the galaxy
\citep[see][]{Tissera2010}. Even though, crudely, the physics of the
formation and evolution of the dark and the stellar halos are the
same, their redshift $z=0$ properties may be drastically different due
to the highly stochastic nature of the stellar halo accretion
\citep[see e.g.][]{Cooper2010}. Indeed, in numerical simulations of
 stellar halo formation, the anisotropy behavior differs markedly
from that of the DM one. Already in the central parts of the Galaxy,
$\beta$ quickly reaches $\sim0.5$ and rises further to $\beta\sim0.8$
at 100 kpc, showing a much smoother monotonic behavior all the way to
the virial radius \citep[see e.g.][]{Abadi2006, Sales2007,Rashkov2013,
  Loebman2017}.
\begin{figure*}
  \centering
  \includegraphics[width=0.98\textwidth]{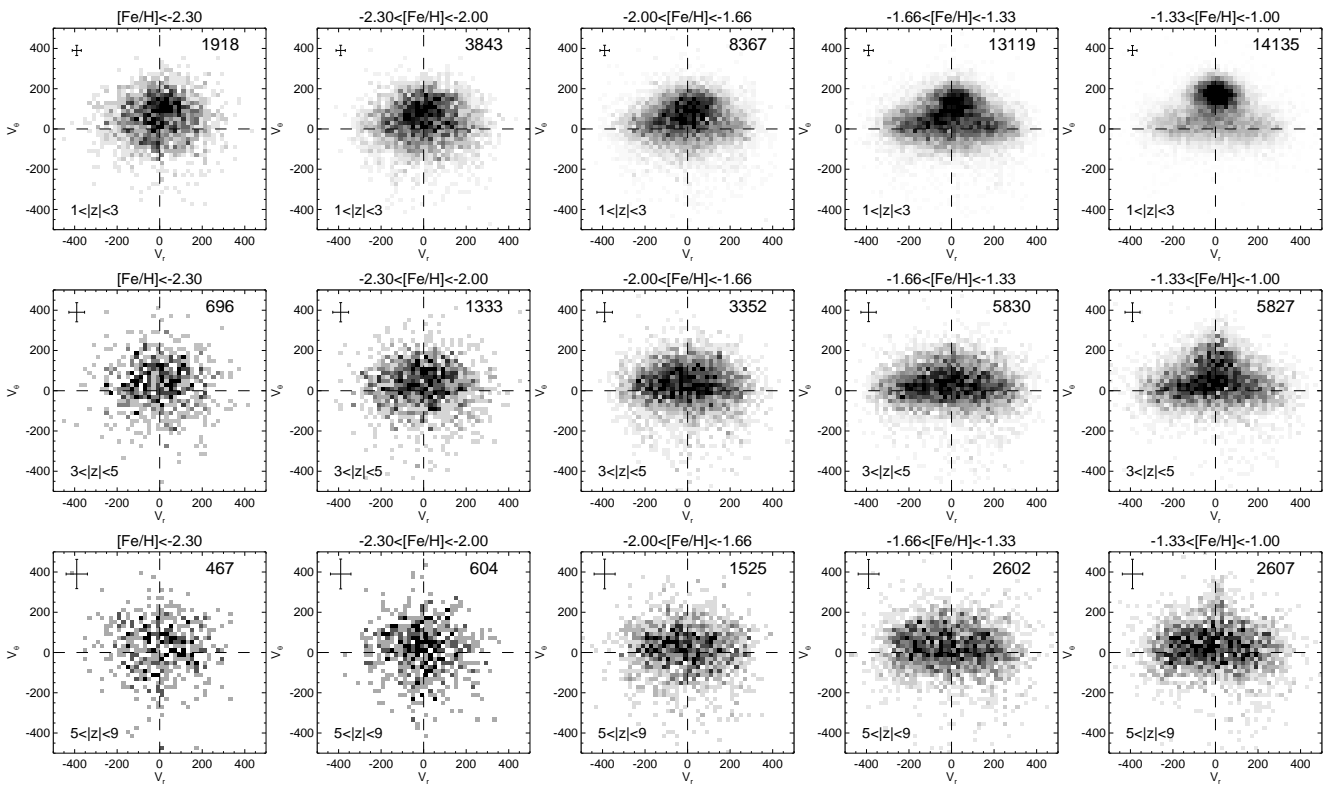}
  \caption[]{Behavior of the velocity components in spherical polar
    coordinates, namely radial $v_r$ and azimuthal $v_{\theta}$, for
    stars in the SDSS-\Gaia\ Main Sequence sample. The dataset is
    divided into metallicity bins, [Fe/H], increasing from left to
    right and Galactic height bins, $|z|$, increasing from top to
    bottom. The size of the median velocity error for each subset is
    shown in the top left corner of each panel and the total number of
    stars in that bin can be found in the top right corner. Pixel size
    is $20\times20$ kms$^{-1}$. The distribution of the metal-rich stars
    near the Galactic plane (top right corner of the panel grid)
    displays two distinct populations: a cold rotating one (the
    thick disc) and a barely rotating, markedly radially anisotropic
    one (the stellar halo). Moving from top to bottom, i.e. to greater heights, the Galactic
    disc contribution quickly diminishes. From right to left, i.e. from metal rich to metal poor, the
    stellar halo's velocity ellipsoid evolves from strongly radial to
    significantly more isotropic.}
   \label{fig:ellipsoid}
\end{figure*}
\begin{figure*}
  \centering
  \includegraphics[width=0.98\textwidth]{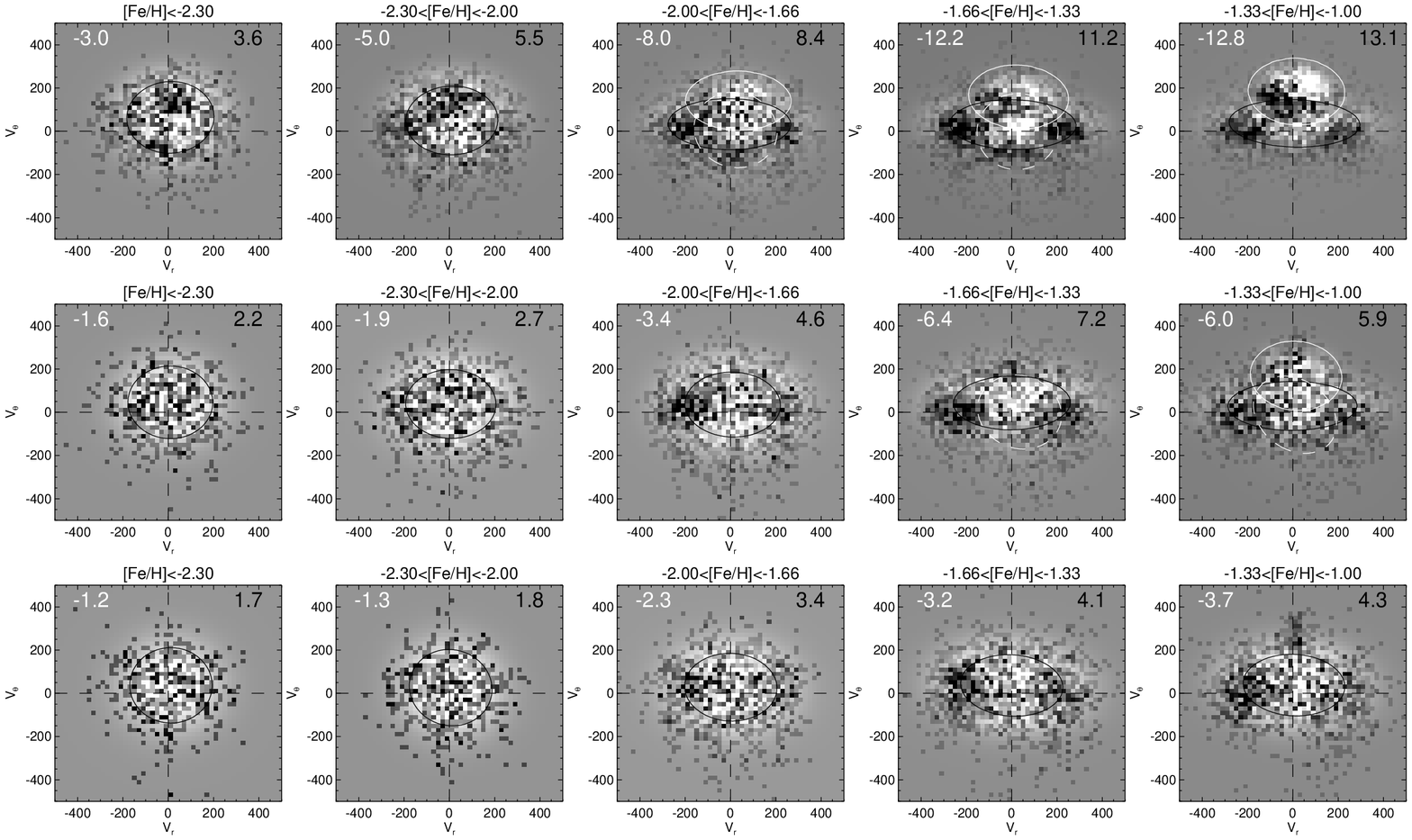}
  \caption[]{Residuals (data-model) of the fit to the distributions
    shown in Figure~\ref{fig:ellipsoid}. Dark (white) corresponds to
    an excess (depletion) of stars in the data compared to the
    model. The number of stars per pixel contributing to the highest
    positive (negative) residuals are shown in the right (left) corner
    of each panel. The Gaussian components used to describe the data
    are also shown: solid white for the disc, solid black for the main
    halo, and white dashed for the additional halo-like component. Note a
    clear pattern of excess-depletion-excess along the radially
    stretched halo component in the right hand side of the grid. The
    appearance of the metal-rich halo residuals clearly indicates that
    the observed velocity distribution is strongly non-Gaussian. Also
    note a small but discernible disc-like residuals in the top-left
    corner of the grid (corresponding to the metal-poor stars near the
    Galactic plane).}
   \label{fig:residual}
\end{figure*}

In this Paper, we explore the behavior of the stellar halo velocity
ellipsoid as a function of metallicity and distance above the Galactic
plane. Our sample is based on the SDSS DR9 spectroscopic dataset and
consists of Main Sequence stars with distances between $<1$ and
$\sim10$ kpc from the Sun. The stars included in the analysis overlap
widely with selections considered by several groups previously
\citep[see][]{Smith2009,Bond2010,Loebman2014,Evans2016}. However,
thanks to the \Gaia\ DR1, the proper motions of these stars are now
measured with an improved accuracy. Section~\ref{sec:data} gives the
details of the stellar sample under consideration, as well as the
model used to characterize the stellar halo behavior. The implications
of our findings and the comparison to the numerical simulations of the
stellar halo formation are presented in Section~\ref{sec:disc}.

\section{Data and Modelling}
\label{sec:data}

\subsection{Main Sequence stars in SDSS-\Gaia}

We select Main Sequence stars from the SDSS DR9 spectroscopic sample
\citep[][]{Ahn2012} by using the following cuts: $|b|>10^{\circ}$,
$A_g<0.5$~mag, $\sigma_{\rm RV}<50$~kms$^{-1}$, S/N$>10$,
$3.5<\log(g)<5$, $0.2<g-r<0.8$, $0.2<g-i<2$, $4500$~K$<T_{\rm
  eff}<8000$ K and $15<r<19.5$. All magnitudes are de-reddened using
dust maps of \citet{SFD}. A total of 192,536 stars survives the above
cuts. We estimate the stars' distances using equations (A2), (A3) and
(A7) in \citet{Ivezic2008}. The proper motions are obtained using a
crossmatch between the SDSS and \Gaia\ catalogs. The SDSS catalog has
been astrometrically recalibrated using the {\texttt GaiaSource}
positions in order to correct both small and large scale systematic
astrometric errors. The resulting SDSS-\Gaia\ proper motion catalog
covers the entirety of the high latitude SDSS sky with baselines
between 6 and 18 years. \citet{Deason2017spin} and \citet{deBoer2018}
assess the quality of the catalog and demonstrate that it suffers
little from systematic biases and clearly outperforms the
SDSS-\Gaia\ catalogs relying on the original SDSS astrometric
solution.  For each star, given the position on the sky and the
distance, the proper motion and the line-of-sight velocity are
converted to Galacto-centric Cartesian velocity components assuming
the Local Standard of Rest of 235 kms$^{-1}$ and the components of the
Solar peculiar motion presented in \citet{LSR}. The uncertainties in
distance, line-of-sight velocity and proper motion are propagated
using Monte-Carlo (MC) sampling to obtain the final uncertainties on
the velocity components in spherical polar coordinates. Accordingly,
for each star, its Galacto-centric coordinates and velocity components
are the median values of the resulting MC distribution, and the
associated ``error-bars'' are the Median Absolute Deviation values
scaled up by a factor of 1.48. To remove the obvious outliers we cull
$0.1\%$ of stars with excessively high speeds. Note that the
(magnitude-independent and metallicity-independent) proper motion
uncertainties are estimated using equation (2) provided in
\citet{Deason2017spin}, \citep[for further discussion also
  see][]{deBoer2018}. Furthermore, neither the selection procedure above
nor the SDSS spectroscopic targeting scheme (with the exception of the
K-giant sample not used here) include restrictions based on the
stellar kinematics. Therefore, we believe that the sample considered
here is kinematically unbiased. Figure~\ref{fig:sample} shows the
density distribution of stars in our sample. A strong spatial bias due
to the SDSS spectroscopic selection is clearly visible in the left
panel, where stars can bee seen reaching $3<R$ (kpc)$<23$ and $|z|<9$
kpc. The right panel gives the distribution of the stellar metallicity
as a function of the Galactic height. A metal-rich, i.e. [Fe/H]$\sim0$
component corresponding to the Galactic disc is apparent at $|z|<1$
kpc. Curiously, a low-height density enhancement is also discernible
at low metallicities, i.e. [Fe/H]$<-1$.

Besides the survey-induced systematics, there may exist other
observational biases that could affect the measurements reported in
this work. For example, a large and non-constant fraction of
unresolved binary objects amongst the stars considered here could
influence the kinematics in the following way. The stellar distances,
as predicted by the equations in \citet{Ivezic2008} are appropriate
for single stars. However, if instead a star is an unresolved binary,
it will appear brighter at a fixed color as compared to the fiducial
relation.  This would in turn lead to an underestimated photometric
distance and therefore underestimated tangential velocity.  Despite
the fact that all stars considered here have spectra, most of the
potential binaries would remain undetected. This is because the
fraction of stars with a clear spectroscopic binary signature
(i.e. doubling of all absorption lines) is expected to be small, due
to i) the low spectral resolution of the SDSS, and 2) because the
binary star period distribution peaks at $10^5$ days
\citep[see][]{Raghavan2010}, thus yielding minuscule line shifts. To
understand the possible impact of the binarity on our results, we have
carried out the following simple test.  We have generated a mock
stellar sample corresponding to an old (12 Gyr) metal-poor ([Fe/H]=-1)
stellar population. The masses of the stars are drawn from the
Chabrier IMF, their distances from $\rho \propto r^{-2}$ radial
density distribution (from 0.1 to 100 kpc) and their velocities from a
Gaussian with a constant velocity dispersion of 100 kms$^{-1}$. Assuming
binary fractions of 0, 0.5 and 0.9 and a uniform mass ratio
distribution \citep[see][]{Raghavan2010}, we calculated the impact of
the binaries on the measured tangential velocity dispersions.
Reassuringly, for the substantial 50\% binary fraction, only a 10\%
lower velocity dispersion is registered. For an unrealistically high
90\% binary fraction, one would underestimate the tangential velocity
dispersion by 20\%. Accordingly, we believe that our velocity
dispersion measurements (see below) are largely insensitive to the
presence of unresolved binary stars in the sample.

\subsection{7D view of the local stellar halo}

Figure~\ref{fig:ellipsoid} shows the evolution of the azimuthal,
$v_{\theta} $, and the radial, $v_r$, velocity components of the stellar
sample defined in the previous subsection as a function of metallicity,
[Fe/H], and height above the Galactic disc, $|z|$. Note that we use the
spherical polar convention in which ${\theta}$ is the azimuthal angle
and $\phi$ is the polar angle. In the Figure, the metallicity of the stars
considered increases from left to right and the Galactic height grows
from top to bottom. In the distributions of the stars closest to the
Galactic plane (top row), two components with different kinematic
properties are clearly discernible. The (thick) disc has a negligible
mean radial velocity, $\bar{v}_r\sim0$ kms$^{-1}$, and a significant rotation,
$\bar{v}_{\theta}\sim200$ kms$^{-1}$. Much hotter in terms of its velocity
dispersion, especially in the $r$ direction, is the stellar halo whose
rotation $\bar{v}_{\theta}$ is extremely weak. As the stars get
progressively metal-poorer (from right to left), the rotating
component fades, and disappears completely when observed at greater heights (middle and
bottom row). Note that as shown in the top left
corner of each panel, the velocity error changes strongly as a
function of the Galactic $|z|$ and for the fixed $|z|$ is
approximately constant across the entire metallicity range. The
evolution of the error with height is driven by the dependence of the
tangential velocity on distance and the distance error on the
photometric uncertainty.

The most notable feature of Figure~\ref{fig:ellipsoid} is the
appearance of the stellar halo velocity ellipsoid. The halo's
distribution is demonstrated to be stretched dramatically in the
radial direction for stars in the high metallicity range, i.e. for
$-1.7 < $[Fe/H]$ < -1$. However, at lower metallicity, it evolves
rapidly to an almost spherical shape. To track the behavior of the
stellar halo's velocity ellipsoid, we model the stellar distributions
shown in Figure~\ref{fig:ellipsoid} with a mixture of multi-variate
Gaussians using the Extreme Deconvolution algorithm described in
\citet{ED}. Given that the number of stars and the complexity of the
overall velocity distribution which evolves quickly as a function of
metallicity and Galactic $|z|$, we use different numbers of Gaussian
components for each sub-sample. The two lowest metallicity bins and
the highest $|z|$ bin are always modelled with one 3D Gaussian
(according to the number of the velocity dimensions in spherical
polars). For all other subsets, in a 3x2 array in the top right corner
of Figure~\ref{fig:ellipsoid} we attempt to fit three 3D Gaussian
components. However, if the resulting component contains less than 3\%
of the total number of stars in a given sub-sample we do not report
its properties. The uncertainties of the model parameters are
estimated using the bootstrap method.

\begin{figure*}
  \centering
  \includegraphics[width=0.98\textwidth]{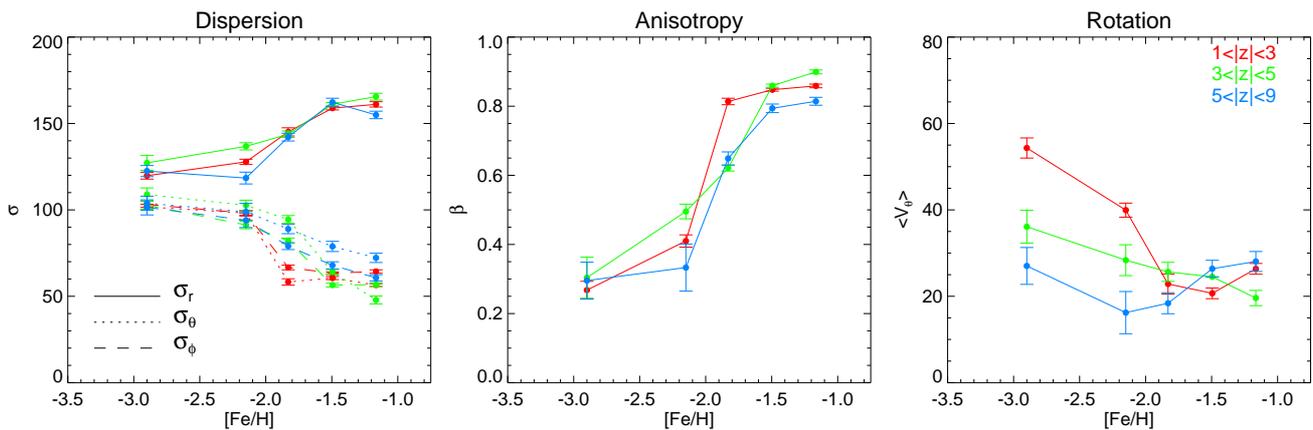}
  \caption[]{Derived properties of the velocity ellipsoid of the main
    stellar halo component as a function of metallicity for three
    different distances from the Galactic plane (indicated by
    color). {\it Left:} Velocity dispersion along each of the three
    dimensions. Note that for [Fe/H]$>-1.7$, the radial velocity
    dispersion ($\sigma_r$, solid lines) is approximately three times
    larger than those in $\theta$ (dotted) and $\phi$ (dashed)
    dimensions. At the metal-poor end, the three velocity dispersions
    come much closer to each other. {\it Middle:} Velocity anisotropy
    $\beta$ of the stellar halo. Note a sharp change in $\beta$ from
    nearly isotropic at the low metallicity to extremely radial for
    the chemically enriched stars. {\it Right:} Amplitude of rotation
    $\bar{v}_{\theta}$. The spin of the metal-rich stellar halo
    sub-population $20<\bar{v}_\theta ($kms$^{-1})<30$ is independent of
    the Galactic height. The amount of rotation in the metal-poor
    stars appears to evolve with $|z|$. However, as explained in the
    main text, this is largely the result of low but non-negligible
    disc contamination and over-simplified model (one Gaussian
    component for the two most metal-poor sub-samples).}
   \label{fig:beta_mean}
\end{figure*}

The results of the multi-Gaussian decomposition of the velocity
distributions are shown in Figure~\ref{fig:residual}. Here, the
significant (i.e. containing $>3\%$ of total stars) model components
are over-plotted on top of the model residuals for each sub-set in the
metallicity and $|z|$ space. White solid lines give the projection of
the disc's velocity ellipsoid, and black solid lines show the main
halo component. In four cases, two in the top and two in the middle
row, a third halo-like Gaussian (white dashed) was required to
describe the data: in these cases its contribution varied between 5\%
and 15\% of the total number of stars in the bin. As displayed in the
Figure, the overall quality of the fits is good, changing from
excellent at the metal-poor end to satisfactory at the metal-rich
extreme. Most metal-rich distributions show clear over-densities of
stars with high radial velocities, both positive and
negative. These high $v_r$ ``lobes'' are most visible in the two right
columns of the Figure. Additionally, some smaller amplitude residuals
are also discernible, e.g. for the stars closest to the disc, in the
most metal-rich bin (top right corner of the grid) and e.g. at
$v_{\theta}\sim150$ kms$^{-1}$ for the stars with lowest metallicity (top left
corner and adjacent).

Figure~\ref{fig:beta_mean} summarizes the evolution of the individual
velocity disperions $\sigma_r, \sigma_{\theta}$ and $\sigma_{\phi}$
(left) and the resulting orbital anisotropy (middle) of the stellar
halo as a function of metallicity for three different Galactic height
ranges. Also shown is the behavior of the rotation $\bar{v}_{\theta}$
of the main halo (right). As hinted in the Figures~\ref{fig:ellipsoid}
and \ref{fig:residual}, $\sigma_r, \sigma_{\theta}, \sigma_{\phi}$ and
$\beta$ are all a strong function of the stellar metallicity.  However, this
dependence does not appear to be gradual. Instead, a sharp
transition in the properties of the stellar halo's ellipsoid can be
seen at [Fe/H]$\sim-1.7$. For more metal-rich halo stars, the
anisotropy is acutely radial, with $\beta$ reaching values of 0.9!
However, the halo traced by the most metal-poor stars is almost
isotropic with $0.2<\beta<0.4$. Note also that while a visual inspection
of Figure~\ref{fig:ellipsoid} registers a noticeable swelling of the
velocity ellipsoid with Galactic $|z|$, this effect is mostly due to
an increase in the velocity error (as shown in the top left corner of
each panel). The final, deconvolved velocity dispersions take each
star's velocity uncertainty into account and show little change in the
velocity ellipsoid shape as a function of $|z|$ (see
Figure~\ref{fig:beta_mean}). In terms of the halo rotation,
interesting yet more subtle trends are apparent. Stars more metal rich
than [Fe/H]$\sim-1.7$ show prograde spin $20 <\bar{v}_{\theta} (\mathrm{kms}^{-1}) <30$
irrespective of the height above the Galactic plane. The rotation of
the metal-poorer stars decreases as a function of $|z|$, from $\sim50$ kms$^{-1}$
near the plane to $<15$ kms$^{-1}$ at $5<|z|({\rm kpc})<9$. Overall, based on the
analysis presented here, the stellar halo appears to be describable,
albeit crudely, with a super-position of two populations with rather
distinct properties.

Ours is not the first claim of the existence of (at least) two
distinct components of the stellar halo \citep[see
  e.g.][]{Chiba2000,Carollo2010}. If the kinematic properties are
measured locally, then it is possible to estimate the 3-D volume
density behavior of each of the halo components \citep[][]{May1986,
  SLZ1990}. Based on the studies above, the metal-richer halo
component was inferred to possess a flatter (with respect to the
Galactic vertical direction) distribution of stars. Note that a
similar argument based on the virial theorem is presented in
\citet{Myeong2018} who analyze a dataset nearly identical to that
presented here and estimate the amount of flattening in each of the
two halo components. Thus, the recently revealed evolution of the
shape of the stellar halo with Galactocentric distance \citep[see
  e.g.][]{Xue2015,Das2016,Iorio2017} could be interpreted as a
change in the contribution from each component. Given that our stellar
sample extends as far as 10 kpc above the disc plane, it may be
possible to track the differences in the density distribution in the
metal-rich and metal-poor sub-populations. Unfortunately, this
calculation is not straightforward to carry out in view of strong
(metallicity and magnitude dependent) selection biases of the SDSS
spectroscopic survey. At the zeroth order - and not accounting for any
selection biases - we do not record any significant changes in the
fraction of stars locked in the metal-richer component (i.e. with
$-1.66<$[Fe/H]$<-1$) with Galactic height: it remains at a level of
$\sim$66\% across all $|z|$ studied.

\section{Discussion and Conclusions}
\label{sec:disc}

This Paper uses the SDSS-\Gaia\ proper motions and thus our dataset is
nearly identical to that of \citet{Myeong2017, Myeong2018}. As shown
by \citet{Deason2017spin} and \citet{deBoer2018}, both random and
systematic proper motion errors in the SDSS-\Gaia\ catalog are
minimized compared to most previously available catalogs of similar
depth. Instead of attempting to identify a sub-sample of halo stars,
we model the entirety of the data with a mixture of multi-variate
Gaussians. This allows us to avoid any obvious selection biases and
extract a set of robust measurements of the stellar halo velocity
ellipsoid. As seen by SDSS and \Gaia, the properties of the local
stellar halo are remarkable. We show that the shape of the halo's
velocity ellipsoid is a strong function of stellar metallicity, where
the more metal-rich portion of halo, i.e. that with $-1.7<$[Fe/H]$<-1$,
exhibits extreme radial anisotropy, namely $0.8<\beta<0.9$ (see
Figure~\ref{fig:beta_mean}). Co-existing with this highly eccentric
and relatively metal-rich component is its metal-poor counterpart with
$-3 <$[Fe/H]$ < -1.7$ and $\beta\sim 0.3$. The velocity ellipsoid
evolves rapidly from mildly radial to highly radial over a very short
metallicity range, changing $\beta$ by $\sim 0.6$ over 0.3 dex. The
two stellar halo components also display slightly different rotational
properties. The more metal-enriched stars show a mild prograde spin of
$\sim25$ kms$^{-1}$ irrespective of Galactic height $z$. The spin of
the metal-poor halo evolves from $\sim 50$ kms$^{-1}$ at $1<|z|(\mathrm{kpc})<3$
, to $\sim20$ kms$^{-1}$ at $5<|z|(\mathrm{kpc})<9$. Note, however, that
while the stellar halo's properties change sharply at
[Fe/H]$\sim-1.7$, we do not claim that the more isotropic component
does not exist at higher metallicities [Fe/H]$>-1.7$, it is just much
weaker (see white dashed contours in Figure~\ref{fig:residual}).

Before we discuss the implications of the measurements presented above
for the genesis of the stellar halo, let us briefly compare the
properties of the velocity ellipsoid deduced here to those in the
literature. Most of the recently published halo anisotropy values are
close to $\beta< 0.7$ \citep[e.g.][]{Smith2009, Bond2010, Posti2017},
significantly lower than $\beta\sim0.9$ reported here. We believe,
however, that there is no obvious disagreement as previous estimates
combine stars across the entire metallicity range and therefore
represent an average over the [Fe/H]-dependent values we measure. In
terms of the halo spin, the amount of prograde rotation at high $|z|$
is consistent with the majority of both the past and recent
studies of the stellar halo \citep[see
  e.g.][]{Chiba2000,Deason2017spin}.

What could be the cause of the dramatic radial anisotropy registered at
high metallicity and what is the nature of the striking bimodality in
the behavior of the stellar halo presented in the previous section?
The two currently available theories for the formation of the stellar
halo in the vicinity of the Galactic plane invoke i) accretion and
disruption of multiple satellites and ii) in-situ formation of a
puffed-up rotating halo via the disc heating. However, it is hard to fathom
how the observed extremely radial velocity anisotropy could be
reconciled with either of these scenarios without any modification. We
are not aware of any in-situ halo model where i) the resulting halo
possesses highly radial anisotropy and ii) the rotation signal would be
as low as measured here, i.e. $\sim25$ kms$^{-1}$ (as judged by the
metal-richer subset of our halo stars). Equally difficult to imagine
is the idea that a prolonged accretion of multiple low-mass Galactic
fragments can yield such high values of $\beta$. Given that the
accreted satellites should show a diversity of orbital properties, it
is more natural to expect a much more isotropic velocity ellipsoid,
perhaps similar to what we observe at low metallicities, but
wildly different from that of the metal-enriched stars, i.e. the bulk
of the stellar halo near the Sun.

Recently, several arguments have been put forward to support the idea
of the halo formation through a single dominant accretion event some
$\sim10$ Gyr ago \citep[see e.g.][]{Deason2013b,unmixing}. In this
scenario, one massive merger provides the lion's share of the halo
stars within $\sim30$ kpc of the Galactic center. At this
characteristic radius \citep[see e.g.][]{Deason2011, Sesar2011}, the
properties of the stellar halo appear to change dramatically thus
leading \citet{Deason2013b} to argue that this transition scale may
correspond to the last apo-centre of the halo's progenitor before
disruption. The hypothesis in which a large portion of the inner
Galaxy's halo is dominated by the stellar debris from a massive
satellite is consistent with the abundance patterns of light elements
in the halo.  The absence of prominent sequences corresponding to
contributions from low-mass systems as well as the iron abundance at
the characteristic ``knee'' of the stellar halo's [$\alpha$/Fe]
distribution \citep[see e.g.][]{Venn2004,Tolstoy2009, Deboer2014}
could all be explained with the (early) accretion of a massive
progenitor. This interpretation is also supported by the study of
\citet{Amorisco2017deposition} who has built a large library of toy
models of idealized merger events in an attempt to provide an atlas of
the halo properties corresponding to different accretion
histories. They confirm the conclusions of \citet{Deason2013b} that
massive satellites are able to sink deeper in the potential well of
the Galaxy due to dynamical friction, thus dominating the inner halo
of the Galaxy. Additionally, \citet{Amorisco2017deposition} show that
the evolution of the orbital properties of the in-falling dwarfs
depends strongly on their mass: the low-mass systems might experience
some mild orbital circularisation, while the orbits of the more
massive systems tend to radialise rather strongly. Therefore, for high
mass-ratio events, due to the pronounced orbital radialisation, the
resulting stellar halo possesses negligible spin at redshift
$z=0$. This is in excellent agreement with the most recent measurement
by \citet{Deason2017spin} and the results presented here.

\begin{figure}
  \centering
  \includegraphics[width=0.48\textwidth]{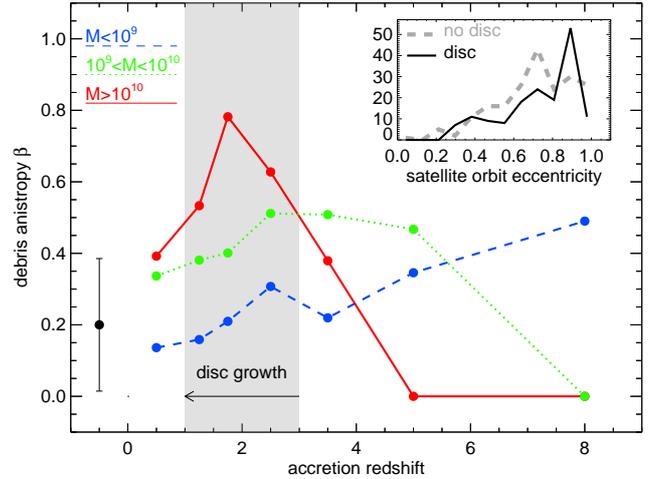}
  \caption[]{Velocity anisotropy of the stellar debris in the Solar
    neighborhood ($4<R(\mathrm{kpc})<25$, roughly matching the range probed by the
    data) as a function of the progenitor's accretion redshift for
    three different sub-halo mass ranges. Each data-point gives the
    median value across the suite of 10 simulations. The highest
    radial anisotropy (for the considered range of Galactocentric
    radii) is attained by the debris from the most massive progenitors
    ($M_{\rm vir}>10^{10} M_{\odot}$, red line) accreted between $z=1$
    and $z=3$, i.e. during the disc growth phase (correspond to 8 to
    11 Gyr of lookback time, grey shaded region). The error-bar on the
    left hand side gives an estimate of the typical scatter in an
    individual (redshift, mass) bin. In the inset, the black solid
    (grey dashed) line shows the distribution of final
    orbital eccentricity for the satellites with $M_{\rm vir}>10^{10}
    M_{\odot}$ and accretion redshift $1<z<3$ with (without) the
    baryonic disc included. Even without the action of the disc, the
    distribution of eccentricities has a peak at $e>0.7$. The presence
    of the disc enhances the orbital radialisation, pushing the peak
    of the distribution to $e>0.9$.}
   \label{fig:sims}
\end{figure}

We seek to verify the above hypothesis using a suite of numerical
simulations which improve on those discussed in the literature so far
\citep[such as those by][]{BJ2005, Amorisco2017deposition}. We
consider 10 cosmological zoom-in simulations of the formation of the
stellar halo around galaxy hosts with masses similar to that of the
Milky Way. These simulations are described in detail in
\citet{Jethwa2018} and \citet{unmixing} and are run with the $N$-body part of
\textsc{gadget-3} which is similar to \textsc{gadget-2} \citep{gadget2}. Our stellar halo realisations
\citep[based on tagging the most bound particles in satellites;][]{delucia_helmi, bailin_2014} are similar to those presented in \citet{BJ2005} and
\citet{Amorisco2017deposition} in the way that they do not attempt to
model the gas-dynamics and the feedback effects. Yet they contain some
of the salient features necessary to understand the stellar halo
emergence, such as the mass function and the accretion time of the
satellite galaxies, as well as the effect of the Galactic disc on the
properties of the Dark Matter halo and the in-falling sub-halos. In
our suite, all 10 simulations are run twice, once without a disc, and
once including the effects of a disc. The disc is represented by a
parametric Miyamoto-Nagai potential \citep[see][]{MN1975}, whose mass
is grown adiabatically from redshift $z=3$ to $z=1$, or in lookback
time, from 11 to 8 Gyr ago. Figure~\ref{fig:sims} shows the orbital
properties of the stellar debris as observed today in the vicinity of
the Sun as a function of the accretion time for three different
progenitor mass ranges. Clearly, the simulated redshift $z=0$ halo
contains a mixture of debris with a wide range of orbital
properties. However, a trend is discernible in which the most massive
satellites (red line) contribute stars on strongly radial
orbits. Moreover, the highest anisotropy values $\beta\sim0.8$ are
recorded for the stellar debris deposited during the phase of the disc
assembly (grey region).

Such synchronicity between the epoch of the disc growth and the
occurrence of massive accretion events is not surprising. Before
sub-halos with virial masses $M_{\rm vir}>10^{10} M_{\odot}$ can be
accreted and destroyed they have to have time to grow. For example,
extrapolating the median halo formation time as shown in Figures 4 and
5 of \citet{Giocoli2007}, the most massive Milky Way satellites would
not have been assembled before redshift $z\sim2$. However, there are
additional factors that may explain the coincidence of the disc
development and the peak in the $\beta$ profile in
Figure~\ref{fig:sims}. The inset in the top-right corner of the Figure
shows the distribution of the orbital eccentricities of the subhaloes
with $M_{\rm vir}>10^{10} M_{\odot}$ accreted at $1<z<3$. Even without
the action of the baryonic disc (dashed grey line), the eccentricity
distribution appears to peak at $e>0.7$, in agreement with the
findings of \citet{Amorisco2017deposition}. Note however that the peak
of the distribution is pushed to even higher values, i.e. $e>0.9$ when
the disc is included (solid black line). Thus, we conclude that the
satellite radialisation at the high mass end is enhanced by the
presence of the growing disc, acting to promote extreme values of the
velocity anisotropy of the inner stellar halo as observed today.

Using the intuition informed by the numerical experiments described
above, let us return to the view of the stellar halo given in
Figures~\ref{fig:ellipsoid} and \ref{fig:residual}. The radically
radial component of the nearby stellar halo is also the one that
contains the most metal-rich halo stars, in agreement with the
mass-metallicity relationship observed in dwarf galaxies
\citep[][]{Kirby2013}. Note however, that in the $(v_r, v_{\theta})$
plane, the density distribution of this sausage-like population is not
exactly Gaussian, as manifested by a pronounced excess of stars with
high positive/negative radial velocity (see
Figure~\ref{fig:residual}). We believe, however, that the
black-white-black pattern of the residuals' distribution in the
right-hand side of the Figure can be equally well explained by the
lack of stars with low radial speeds. This interpretation may be more
appropriate if the Sun is located somewhere in the middle of a giant
cloud of stars on highly radial orbits (with their peri-centres at
small Galactic radii and the apo-centres not far from the break radius
at $\sim$30 kpc). Limited by the available data, our view of the halo
lacks stars near the turning points - apos and peris - where the
radial motion is the slowest. We have validated this idea using the
numerical simulations described above. Most importantly, given that
due to the selection effect described above, the velocity distribution
is noticeably non-Gaussian, the (already high) anisotropy value for
the metal-rich component may actually be an under-estimate.

Notwithstanding the strong radialisation of the progenitor's orbit, it
is unlikely that it would lose all of its angular momentum. In this
regard, the low-amplitude yet clearly detectable spin of the
metal-rich halo component could simply be the relic of the orbital
angular momentum of the parent dwarf before dissolution. This
interpretation also agrees with the observed constancy of the rotation
amplitude of the metal-rich halo with the height above the Galactic
plane. Given the range of vertical distances probed here, any
significant change in the amount of rotation would imply that the
scale-height of the halo component considered is not dissimilar to
that of the thick disc's. This can be contrasted to the behavior of
the metal-poor halo sub-population. The amount of rotation at the
metal-poor end is inversely proportional to the height above the
plane. The strongest signal of $>50$ kms$^{-1}$ is detected near the
Galactic plane. As the two top left panels of
Figure~\ref{fig:residual} demonstrate, the single-Gaussian model
produces noticeable residuals at $v_{\theta}\sim150$ kms$^{-1}$. Given
the appearance of the model residuals for the lowest height sub-sample
(top left panels), we conjecture that the halo spin is probably an
over-estimate due to the presence of a non-negligible number of
non-halo stars in apparent prograde rotation. Interestingly, at
metallicities [Fe/H]$<-1.3$ we recover a small but statistically
significant positive radial velocity for stars contributing to this
disc-like population. This is consistent with the analysis of
\citet{Myeong2018} who also show that the excess of rotating
metal-poor stars with $0<v_r(\mathrm{kms}^{-1})<20$ is a strong function of
Galacto-centric radius. Thus, it is unlikely that this signal is
provided by a (thick) disc alone. Instead we conjecture that this
could plausibly be a combination of an extremely metal-poor disc and
stars trapped in a resonance with a bar, a feature known as the
Hercules stream \citep[see][]{Dehnen2000,Antoja2014,Hunt2018}. At
larger distances from the Galactic plane, the contribution of the disc
at the metal-poor end sharply subsides and the velocity distribution
can be adequately described with a single Gaussian, yielding spin
values comparable (or sometimes slightly lower) to those observed at
the metal-rich end.

An alternative explanation of the strong radial anisotropy observed in
the stellar halo may be provided by the ideas of \citet{ELS}, i.e. by
invoking a non-adiabatic mass growth which would result in an increase
of orbital eccentricities in the halo. However, this dramatic
contraction of the young Galaxy would affect all stars situated in the
Milky Way at that time, irrespective of their metallicity. This can be
contrasted with the measurements reported here: at the metal-poor end,
a much lower velocity anisotropy is observed, $0.2<\beta<0.4$. This
nearly isotropic velocity ellipsoid can still be reconciled with the
hypothesis of the collapse-induced radialisation if the metal-poor
stars were either accreted after the prolific growth phase or stayed
sufficiently far away from the portions of the Galaxy undergoing rapid
transformations. However, as Figure~\ref{fig:sims} demonstrates, there
may be a simpler explanation of the dependence of the velocity
ellipsoid on the stellar metallicity. According to the Figure,
moderate radial anisotropy appears consistent with the stellar debris
contributed by the low-mass objects, irrespective of the accretion
redshift. Thus, the Cosmological zoom-in simulations paint a picture
in which the stellar halo dichotomy emerges naturally due to a
correlation between the orbital properties of the accreted dwarfs and
their masses. 

\section*{Acknowledgments}

The research leading to these results has received funding from the
European Research Council under the European Union's Seventh Framework
Programme (FP/2007-2013) / ERC Grant Agreement n. 308024.  A.D. is
supported by a Royal Society University Research Fellowship.
A.D. also acknowledges support from the STFC grant ST/P000451/1. NWE
thanks the Center for Computational Astrophysics for hospitality
during a working visit.

\bibliography{references}

\label{lastpage}

\end{document}